\documentclass[prd, twocolumn, nofootinbib, floatfix]{revtex4-1}

\usepackage{amsmath}
\usepackage{graphicx}
\usepackage{dcolumn}
\usepackage{bm}
\usepackage{epsfig}
\usepackage{amssymb,latexsym,mathrsfs}
\usepackage{graphicx}
\usepackage{color}
\usepackage{hyperref}

\hypersetup{
    colorlinks=true,
    linkcolor=red,
    citecolor=blue,
} 

\newcommand{\be}{\begin{equation}}
\newcommand{\ee}{\end{equation}}
\newcommand{\bs}{\begin{split}} 
\newcommand{\bea}{\begin{eqnarray}}
\newcommand{\eea}{\end{eqnarray}}

\begin{document}

\title{Robust Strong Lensing Time Delay Estimation} 
\author{Alireza Hojjati$^1$, Alex G.\ Kim$^{2}$, Eric V.\ Linder$^{1,2,3}$} 
\affiliation{$^1$Institute for the Early Universe WCU, Ewha Womans 
University, Seoul 120-750, Korea\\ 
$^2$Lawrence Berkeley National Laboratory, Berkeley, CA 94720, USA\\ 
$^3$Berkeley Center for Cosmological Physics, 
University of California, Berkeley, CA 94720, USA}

\begin{abstract}
Strong gravitational lensing of time variable sources such as quasars and 
supernovae creates observable time delays between the multiple images.
Time delays can provide a powerful cosmographic probe through 
the ``time delay distance'' involving the ratio of lens, source, 
and lens-source distances.
However, lightcurves of lensed images have measurement gaps, noise, 
systematics 
such as microlensing from substructure along an image line of sight, and 
no a priori functional model, making robust time delay 
estimation 
challenging.  Using Gaussian process techniques, we demonstrate success in 
accurate blind reconstruction of time delays and reduction in uncertainties 
for real data. 
\end{abstract} 

\date{\today} 

\maketitle

%%%%%%%%%%%%%%%%%%%%%%%%%%%%%%%%%%%%%%%%%%%%%%%%%%%%%%%%%%%%%%%%%%%%%%%%
\section{Introduction} 

Multiple images of a single source are dramatic evidence for the effect of 
gravity, specifically general relativity, on light.  This strong 
gravitational lensing not only splits the images, but magnifies or 
demagnifies the source flux and induces time delays between the images.  
The time delays arise from both the geometric path differences along the 
various lines of sight and the gravitational potential differences traversed 
by the photons. 

When the source is variable, such as from a quasar or supernova, the 
time delays in the flux of one image relative to another can be observed. 
With careful modeling of the lens mass distribution, and measurement of 
the angular positions of the images, the geometric factors of distances 
between observer and lens, observer and source, and lens and source can 
be extracted as a ratio called the time delay distance.  Recent advances 
in lens modeling \cite{oguri07,suyu0804} and careful, long term flux 
monitoring programs such as CosmoGrail \cite{cosmograil} (also see 
\cite{suyu0910,fadely}) have matured strong lensing 
time delays to an incipient cosmographic probe. 

This prospect is exciting for several reasons.  Since time delays over 
cosmological distances are sensitive not just to the overall scale, or 
Hubble constant, but the cosmic energy density and its evolution 
with redshift, one can constrain (combinations of) the matter and dark 
energy densities and dark energy equation of state.  Moreover, the time 
delay distance acts fundamentally differently from luminosity and angular 
distances measured by calibrated standard candles such as Type Ia supernovae 
and rulers such as baryon acoustic oscillations.  Hence it has distinct 
covariances among cosmological parameters and can be powerful in 
complementarity with the standard distance probes \cite{lin04,lin11}.  
Finally, despite the lens mass modeling, strong lensing time delays are 
a geometric probe and are tied only to the late universe, unique except for 
supernovae (but with different systematics and covariances) among all 
cosmological probes. 

Here we address one important element of the use of lensing time delays: 
accurate estimation of the actual time delays.  While great progress has 
been made in recent years (see, e.g., \cite{tewes,12086009,12086010}), 
in large part due to 
heroic observing programs and improved data sets, this is not a solved 
problem.  Mathematically, one can consider it as reconstructing a shift 
between multiple noisy, irregularly sampled, differentially amplified 
data streams.  We apply a special combination of Gaussian process statistics 
to this task.  Such a concept for strong lensing dates back to \cite{press} 
and more recently has been shown to have reasonable success \cite{tewes}; 
we introduce several new features that exhibit noticeable improvement in the 
state of art. 

Section~\ref{sec:challenge} outlines the challenge of reconstructing the 
time delays from realistic data complete with systematics such as 
microlensing.  The Gaussian process methodology is described in 
Sec.~\ref{sec:gp}, introducing the various correlation function terms 
and accounting for systematics.  We test the method against blinded mock 
data, and real data from the literature, in Sec.~\ref{sec:test}, and 
conclude in Sec.~\ref{sec:concl}.

%%%%%%%%%%%%%%%%%%%%%%%%%%%%%%%%%%%%%%%%%%%%%%%%%%%% 
\section{Time Delayed Lightcurves} \label{sec:challenge} 

Fluxes received from an image at several times define a lightcurve, 
but the name is misleading since the data are not continuous but discrete, 
and the observations are often irregular and sparse and have measurement 
uncertainties.  The best 
monitoring frequency may be every day or two, while long gaps of a few 
months occur due to seasonal visibility of regions of the sky from a 
single telescope.  The cadence is often irregular, though ongoing wide 
area surveys such as Dark Energy Survey (DES \cite{des}), Kilodegree 
Survey (KIDS \cite{kids}), and PanSTARRS \cite{panstarrs}, and in the 
future LSST \cite{lsst}, may have regular observations with 
periods of several days. 

Apart from the sparseness, the data has photometric measurement noise. 
Most current observations come from small (1 meter) telescopes, and 
atmosphere, telescope, and detector noise all contribute.  With wide 
field surveys, hundreds to thousands of time delay systems may be found, 
enabling choice of the cleanest for use as time delay distance probes.  
Since to obtain a time delay distance one must have a robust model of the 
lens mass distribution, galaxy lenses are preferred over cluster lenses 
due to less complex modeling.  Depending on lens mass and geometry this 
implies time delays in the range of a few to hundred days in general. 

Comparing lightcurves from different images involves some form of 
cross-correlation, looking for the time delay between them.  Straightforward 
cross-correlation techniques tend not to work well due to the noisiness 
and sparseness of the data, and extrinsic contributions (see, e.g., 
\cite{koopmans}).  
Instead of comparing noisy data with noisy data, regression techniques 
attempt to reconstruct the underlying true source variation and 
compare the image measurements to that.  We employ Gaussian processes 
(GP) as the regression technique.  See \cite{kimgp} for an example of 
its application to (non-lensed) supernova lightcurves. 

In addition to measurement difficulties, astrophysical systematics 
contribute to the challenge of time delay estimation.  Further time 
variations arise from microlensing caused by passage of substructure 
near to the line of sight.  This affects images independently, breaking 
the (delayed) coherence between them, and can occur on all time scales.  
Short variations just add noise but long term variations disrupt the 
relation between the lightcurves for large portions of the data set and 
so can cause misestimation of the time delay.  These long term variations 
are moderately smooth and some previous work has used low order polynomials 
or splines to represent them; we instead allow the data to determine 
their time scale. 

\begin{figure}[htbp!]
\includegraphics[width=\columnwidth]{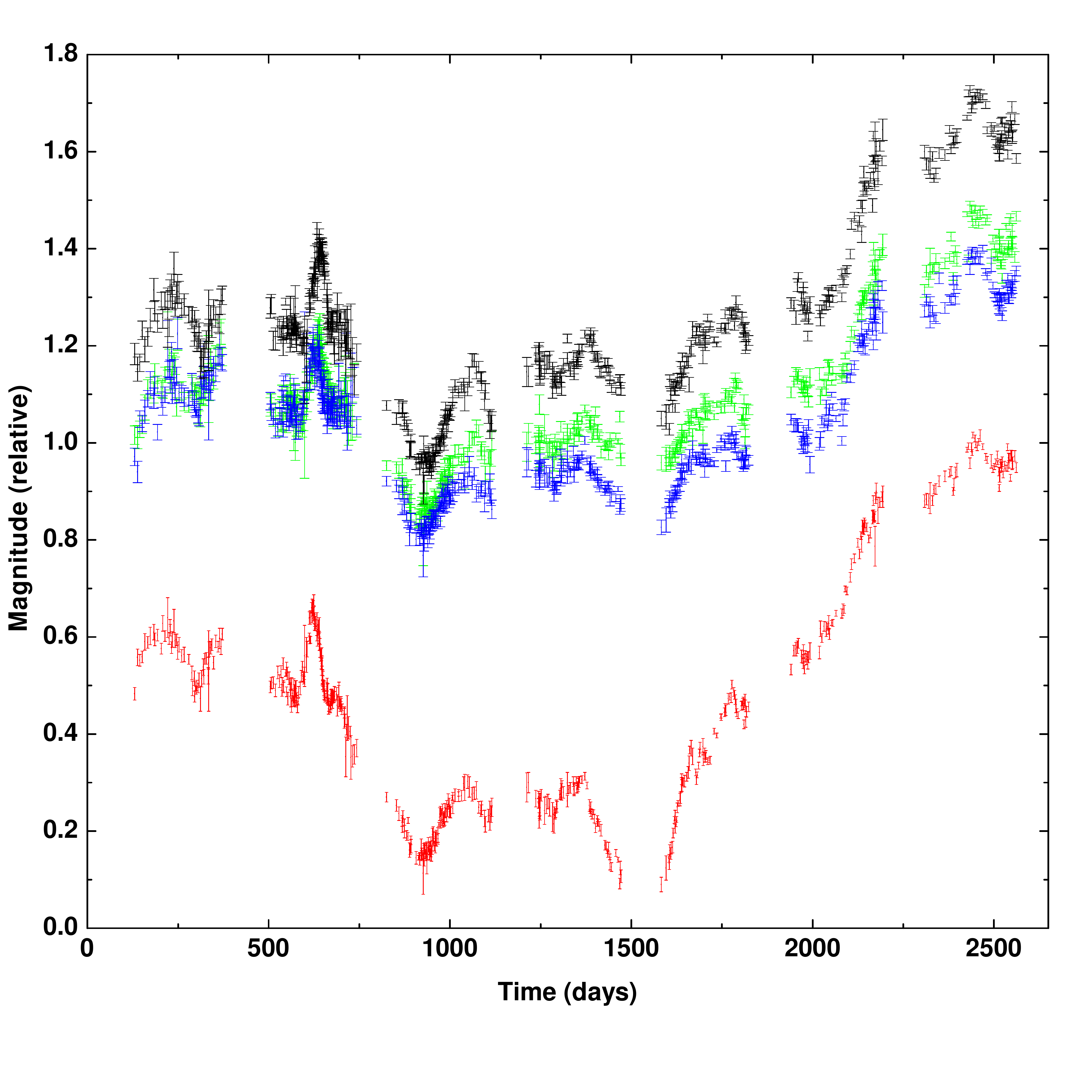}
\caption{Magnitudes (log flux) of four images of the quasar HE 0435-1223 
are plotted vs time, with an arbitrary overall zeropoint. 
}
\label{fig:lche}
\end{figure}

Thus we have three elements entering into the light curves: the intrinsic 
variation that we want to measure, the observational noise, and the 
astrophysical microlensing systematic (in fact our formalism would allow 
multiple versions of the last two).  
The challenge of robust time delay estimation is to reconstruct phase 
shifts of a source with unknown intrinsic flux variation, for images with 
independent microlensing magnifications along their lines of sight, 
using noisy data with irregular temporal sampling.  
Figure~\ref{fig:lche} shows an example 
of real lightcurves from four images of quasar HE 0435-1223 measured by 
CosmoGrail \cite{grailweb}.  Conventionally observations are reported in 
magnitudes (logarithmic flux units).

%%%%%%%%%%%%%%%%%%%%%%%%%%%

%%%%%%%%%%%%%%%%%%%%%%%%%%%%%%%%%%%%%%%%%%%%%%%%%%%% 
\section{Time Delay Estimation} \label{sec:gp} 

For reconstructing an intrinsic function from isolated, noisy data points, 
Gaussian processes offer a robust, substantially model independent 
statistical method with well defined error characterization.  See \cite{gpml} 
for a thorough discussion of GP from 
a statistical point of view.  The basic idea is that the function is not 
parametrized, but rather the data are fit to a whole family of possible 
curves, given by a Gaussian distribution with a mean function and a 
covariance kernel between points.  

The key choices are the form of the mean function (which ideally does not 
affect the final fit but in practice a poor mean function can lead to 
difficulties) and the covariance kernel, together with any hyperparameters 
used in those functions.  There is a single GP representing the 
true source underlying all of the images plus the microlensing of a 
reference image.  For a mean function we adopt a constant function, then 
allow hyperparameters for magnifications relative to the reference image.  
We try different reference images to test for robustness. 

For the covariance kernel we investigate three possibilities.  
A damped random walk (DRW) is often adopted to model 
the intrinsic quasar light curve \cite{drw1,drw2,drw3,drw4}.  
While we are here focused on extracting accurate time delays, 
not modeling the quasar per se, it is natural to try the DRW kernel 
\be 
k(t_i,t_j)=\sigma^2\,e^{-|t_i-t_j|/l} \ , 
\ee 
where $t_i$ and $t_j$ are measurement times, the hyperparameter 
$\sigma$ adjusts the amplitude of the kernel and $l$ functions as a 
correlation length.  

Another choice is a Matern function with index $3/2$, 
\be 
k(t_i,t_j)=\sigma^2 \left(1+\frac{|t_i-t_j|\sqrt{3}}{l}\right)\, 
e^{-|t_i-t_j|\sqrt{3}/l} \ . 
\ee 
The Matern function is commonly used in statistics 
\cite{gpml} and allows for 
greater roughness in the variation than another common choice, the 
squared exponential or Gaussian, 
\be 
k(t_i,t_j)=\sigma^2\, e^{-(t_i-t_j)^2/(2l^2)} \ . 
\ee 
We will compare the results for these three kernels to give extra 
crosschecks on the results; generally we find that DRW works best, 
once guided by an initial Matern run. 

We include measurement noise and an additional nugget term 
$\sigma_n^2\delta_{ij}$, which acts as a zero lag dispersion, e.g.\ 
an empirical term for misestimated measurement noise or finite realization 
scatter. 
This is distinct from the GP 
amplitude $\sigma$ in that $\sigma$ accounts for the global variations of the 
kernel whereas the nugget term $\sigma_n$ accounts for the independent 
dispersion of the individual data points around the predicted GP value.  

The microlensing systematic has been attempted to be addressed in the 
literature by multiplying the lightcurves by a quadratic polynomial or 
a cubic spline over short time spans or within an observing season.  This 
restricts the allowed variations and has the potential to lead to bias 
in the reconstructed time delays or simply a failed fit.  We remain 
within the GP framework, which does not impose a specific model or timescale 
for the microlensing, and account for the microlensing with a GP 
for each image (other than the reference one) 
with zero mean function and a squared exponential kernel of common amplitudes  
$\sigma_\mu^2$ and correlation lengths $l_\mu$.  To separate the microlensing 
GP from the quasar GP, we require a long correlation length $l_\mu$ (systems 
with the microlensing timescale comparable to the intrinsic variations are 
not useful for time delay measurement). 
We have investigated various choices of priors, for example 
$\pi(l_\mu)>50$ days, $\pi(l_\mu)>$\ season, or $\pi(l_\mu)>3l$; all give 
equivalent results.  

We emphasize that neither the intrinsic quasar lightcurve nor the 
microlensing actually have to be (and may not be) true GPs in themselves; 
all we want to test is whether robust time delays can be estimated from 
this approach.  

In summary, the lightcurve predictions for our full GP regression take 
the form 
\bea 
\vec y_1&\sim& GP_Q(\vec\theta_{\rm Qhp};t-t_1)\\ 
\vec y_2&\sim& GP_Q(\vec\theta_{\rm Qhp};t-t_2)+GP_{\mu 2}(\vec\theta_{\mu\rm hp})+\Delta m_2\\ 
\vec y_3&\sim& GP_Q(\vec\theta_{\rm Qhp};t-t_3)+GP_{\mu 3}(\vec\theta_{\mu\rm hp})+\Delta m_3 
\eea 
and so forth for each image, where $\vec\theta_{\rm Qhp}$ is the 
hyperparameter vector for the quasar GP, $\vec\theta_{\mu\rm hp}$ is for 
the microlensing GP, and $\Delta m$ represents the magnification relative 
to the reference image 1. 

The GP likelihood is \cite{gpml} 
\begin{equation} 
2 \ln p(Y|\vec\theta)= - Y^T K^{-1} Y  - \ln |K| - N_d\, \ln 2\pi ,
\label{GP-likelihood} 
\end{equation} 
where $Y$ is the vector of magnitude data, with $N_d$ the total number of 
data points, $\vec\theta$ represents the fit parameters, e.g.\ time delays, 
and $K$ is the full kernel (the sum of the quasar GP, microlensing GP, 
measurement noise, and nugget) with $|K|$ being its determinant.  
The likelihood is maximized for the most likely values 
of the time delays and magnifications, which we find using 
the function minimizer routine Minuit \cite{minuit} and have validated using 
a Monte Carlo analysis. 

In principle, we can combine all lightcurves at once, compare two at a time, 
or any number of lightcurves.  Simultaneous analysis of more than two curves 
allows a consistency check in the form of the triangle equality, e.g.\ 
$\Delta t_{AC}=\Delta t_{AB}+\Delta t_{BC}$, and is our baseline approach.  
Using more lightcurves also has the advantage of the leverage of more images 
on simultaneously constraining the underlying source light curve.  
Analysis using just a pair has fewer hyperparameters and may 
deliver smaller statistical errors, but at the risk of bias.  We carry out 
crosschecks by trying different numbers of lightcurves in the analysis, 
finding that the results from the pair analyses can provide useful initial 
conditions to the simultaneous fit.  
One can also use portions of data, such as selected observation seasons, 
to cross check the consistency of the results or to reduce the impact of 
microlensing as has been done in the literature before.  We find the 
results from our approach to be robust to the number of data points used 
in the analysis. 

In summary, when fitting $N$ lightcurves we have the $N-1$ time delay 
parameters that are our goal, the $N-1$ magnifications $\Delta m$, and 
the hyperparameters $\sigma^2$, $\sigma_n^2$, $\sigma_\mu^2$, $l$, $l_\mu$.

%%%%%%%%%%%%%%%%%%%%%%%%%%%%%%%%%%%%%%%%%%%%%%%%%%%% 
\section{Tests and Results} \label{sec:test} 

\subsection{Blind mock data} \label{sec:blind} 

To test the accuracy and robustness of the method we initially created 
blinded mock data sets.  To preserve realistic sampling and data quality, 
one author took lightcurve data from one image of quasar HE 0435-1223, 
realized three new lightcurves using 
random Gaussian distributions with mean zero and 
standard deviation equal to the data errors, 
and shifted each of the resulting lightcurves vertically by 
various magnifications and horizontally by time delays.  The shifted 
data were then resampled onto the original time sampling using linear 
interpolation.  
Another author, unaware of the simulated time delay and magnification 
values, was given the final data points with error bars and carried 
out the GP fit. 

The results are shown in Table~\ref{tab:blind}, with the true values of 
15.0 and 25.0 day delays recovered within the 68\% confidence level by 
each of the three covariance functions.  Several 
other tests with different time delays had similar results.  
The DRW 
kernel gives results that are significantly more precise, but due to its 
allowance of high level of variations we find that it works best when 
we first run a GP with a Matern kernel, and use that result as a prior with 
10 times the Matern time delay uncertainties when running DRW.  

We find that the magnification 
and nugget terms are both important to include.  Time delays are 
also tested for robustness by choosing different reference curves and 
different multiplicities (i.e.\ fitting for the AB time delay in isolation, 
or simultaneously fitting the GP to more than two lightcurves).  Quoted 
values reflect the central values and uncertainties from the configuration 
that has the best reduced $\chi^2$ and the smallest errors.  These 
uncertainties are marginalized over all the other parameters and 
hyperparameters; the distributions are sufficiently Gaussian that the 
68\% CL error bars are symmetric.

%%%%%%%%%%%%%%%%%%%%%% 
%\begin{table*}[!htb]
\begin{table}[!htb]
\begin{tabular}{|c|c|c|c|} 
\hline
Kernel& $\Delta t_{AB}$ & $\Delta t_{AC}$ & $\Delta t_{BC}$\\ 
\hline 
DRW  & $ 14.94 \pm 0.14$ & $24.99 \pm 0.09$ & $10.0 \pm 0.2$\\
Matern & $14.3 \pm 0.8$ & $25.1 \pm 0.9$ & $10.8 \pm 0.9$ \\ 
Sq Exp & $13.9 \pm 1.3$ & $25.8 \pm 1.4$ & $10.6 \pm 0.7$\\
\hline 
\end{tabular}
\caption{Blind analysis of time delays works for DRW, Matern, and squared 
exponential GPs.  The input to the 
simulation had $\Delta t_{AB}=15.0$ days, $\Delta t_{AC}=25.0$ days. 
}
\label{tab:blind} 
%\end{table*} 
\end{table}

\begin{figure}[htbp!]
\includegraphics[width=\columnwidth]{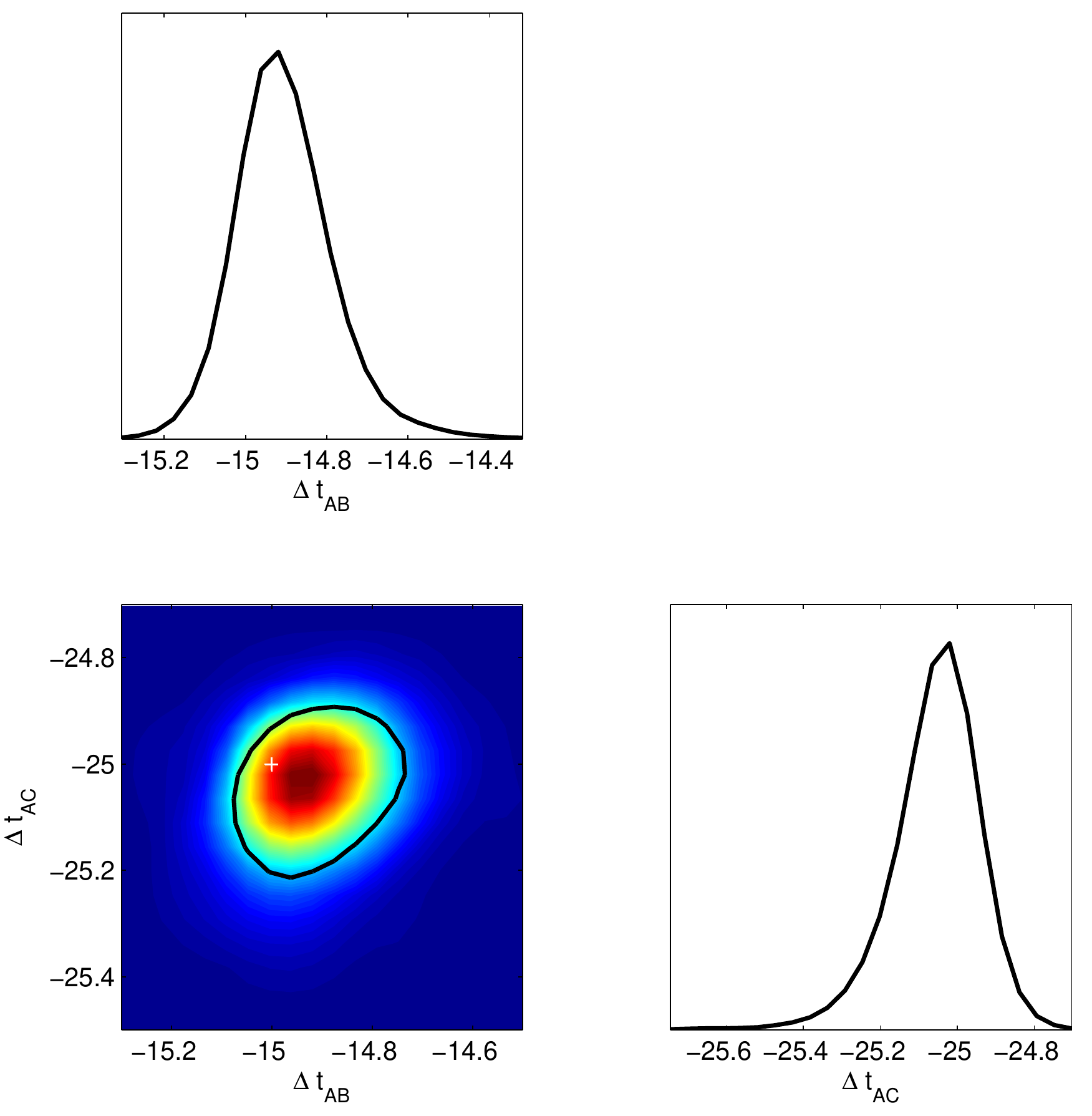}
\caption{Marginalized 1D and 2D likelihood contours are illustrated for 
the two time delays in the mock data case. The fiducial value is marked 
with a white plus sign. 
}
\label{fig:likeli}
\end{figure}

Figure~\ref{fig:likeli} shows the 1D and 2D joint likelihood contours for 
the time delay parameters in the mock data case using the DRW GP.  
As a comparison, these 
results are obtained using CosmoMC \cite{cosmomc} as a generic Monte Carlo 
sampler, and are wholly consistent with the Minuit results.  
For all the parameters 
and hyperparameters we impose a very wide flat prior and let data decide 
their values. The only constraint is on the microlensing correlation length, 
which as discussed should not be too small and hence mix with the actual 
correlation length of the GP kernel.

%%%%%%%%%%%%%%%%%%%%%%%%%%%%%%%

A larger and more sophisticated series of data challenges is forthcoming 
as part of the LSST Dark Energy Science Collaboration strong lensing 
working group.  This will provide large, sophisticated mock data sets and 
an excellent opportunity for testing further development of robust 
time delay estimation.

%%%%%%%%%%%%%%%%%%%%%%%%%%%%%%%%%%%%%%%%%%%%%%%%%%%%%%%%%%% 
\subsection{Actual data} \label{sec:actual} 

The second part of testing the GP method involves 
using public data sets from CosmoGrail and other literature sources 
\cite{Vuissoz:2008qa,cosmograil,Kochanek:2005ge,Fassnacht:2002df,suyu0910} 
as inputs for time delay estimation.  These results can then be compared 
to the literature results obtained using a variety of different methods. 

We use the two sets of CosmoGrail lightcurves publicly available at 
\cite{grailweb}, for quasars HE 0435-1223 and WFI J2033-4723, and the 
radio lightcurves of quasar B1608+656, courtesy of Chris Fassnacht.  
Table~\ref{tab:actual} compares the results we obtain from our GP analysis 
using the DRW and Matern kernels with those published in the literature. 
We also have tested the square exponential kernel but this gives weaker 
uncertainties. 
The values from our analysis and the literature are consistent with each 
other, with the GP analysis tending 
to have smaller uncertainties.  Note the true values of the time delays are 
not known, but the consistency offers an indication of robustness.

%%%%%%%%%%%%%%%%%%%%%% 
\begin{table*}[!htb]
\begin{tabular}{|c|c|c|c|c|c|c|} 
\hline 
Kernel& $\Delta t_{AB}$ & $\Delta t_{AC}$ & $\Delta t_{AD}$ & 
$\Delta t_{BC}$ & $\Delta t_{BD}$ & $\Delta t_{CD}$ \\ 
\hline 
HE 0435-1223 GP-DRW &     $-9.5  \pm 0.3$ & $-1.9\pm0.4$ & $-15.6\pm0.3$& $8.1 \pm 0.3$ & $-6.0 \pm 0.3$& $-13.6\pm0.4$\\ 
HE 0435-1223 GP-Mat &     $-9.6  \pm 1.1$ & $-1.5 \pm 1.1$ & $-14.0 \pm 0.9$ &  $8.1 \pm 1.1$ & $-5.0 \pm 1.1 $ & $-12.3 \pm 1.1 $\\ 
HE 0435-1223 Lit(1) \cite{cosmograil}&  $-8.4 \pm 2.1$ & $-0.6 \pm 2.3$ & $-14.9 \pm 2.1$ &  $7.8 \pm 0.8$ & $-6.5 \pm 0.7$ & $-14.3 \pm 0.8$  \\ 
HE 0435-1223 Lit(2) \cite{Kochanek:2005ge}& $-8.8 \pm 2.4$ & $-2.0 \pm 2.7$ & $-14.7 \pm 2.0$ &  $6.8 \pm 2.7$ & $-5.9 \pm 1.7$ & $-12.7 \pm 2.5$  \\ 
\hline 
WFI J2033-4723 GP-DRW  & $35.1 \pm 0.9 $ & $- 24.9 \pm 0.4 $ & -- & $-59.2 \pm 2.1 $ & -- & -- \\ 
WFI J2033-4723 GP-Mat  & $36.0 \pm 1.5 $ & $- 26.3 \pm 1.7 $& -- & $-62.0 \pm 2.3 $ & -- & -- \\ 
WFI J2033-4723 Lit \cite{Vuissoz:2008qa}& $35.5 \pm 1.4 $ & -27.1 +4.1/-2.3  & -- & ? & -- & --\\ 
\hline 
B1608+656 GP-DRW  & $31.8 \pm 2.4 $  & $ -1.3 \pm 1.5$ & $ -51.0 \pm 6.2 $ & $ -33.1 \pm 2.7$& $ -72.0 \pm 4.5$ & $ -43.1 \pm 3.6$ \\ 
B1608+656 GP-Mat  & $31.7 \pm 2.1 $  & $-2.4 \pm 2.2$ & $ -50.4 \pm 6.9 $ & $-35.0 \pm 4.0$& $-77.5 \pm 7.1$ & $-44.4 \pm 5.4$ \\ 
B1608+656 Lit \cite{Fassnacht:2002df}& 31.5 +2.0/-1.0 & ? & ? & $-36.0 \pm 1.5$&  -77.0 +2.0/-1.0 & ?\\ 
\hline 
\end{tabular}
\caption{Time delay estimations are compared between our GP analysis 
and values in the literature using different reconstruction methods. 
A question mark represents time delay estimates not provided by the 
literature, a dash indicates there is no fourth image. 
}
\label{tab:actual} 
\end{table*}

The GP analysis not only estimates the time delays, a key input for 
cosmography through time delay distances, but provides information on 
the intrinsic quasar variability, the variations around the best fit 
GP lightcurve, and the microlensing systematics through the hyperparameters 
such as the correlation lengths, GP amplitudes, and nugget. 

We find that there is no significant correlation between the parameters.  
The nugget term is usually important and has a value 
comparable to the errors on the data points.  We also find that including 
the microlensing term is useful even when there is no significant 
microlensing in the system. 

The quasar HE 0435-1223 (Fig.~\ref{fig:lche}) has a long observation period 
with distinct features in the intrinsic variability, making it fairly 
straightforward to compute the time delays.  The bottom curve has 
significant microlensing variation which leads to large microlensing 
amplitude $\sigma_{\mu}$. The microlensing correlation length ($\sim 700$ 
days) is completely separated from the quasar GP correlation length 
($\sim 100$ days). There is strong agreement between our results, those of 
Literature 1 \cite{cosmograil} that uses only the first two observation 
seasons, and those of Literature 2 \cite{Kochanek:2005ge}.  Our uncertainties 
are smaller by a factor of several. 

The quasar WFI J2033-4723 has a relatively shorter observation time but 
distinct features in the light curves. There is no significant long-range 
microlensing and hence $\sigma_{\mu}$ is very small indicating that including 
microlensing terms may not be necessary (but this is not known a priori).  
Again, despite using several hyperparameters, our marginalized uncertainties 
are smaller than the results from \cite{Vuissoz:2008qa}. 

\begin{figure}[htbp!]
\includegraphics[width=\columnwidth]{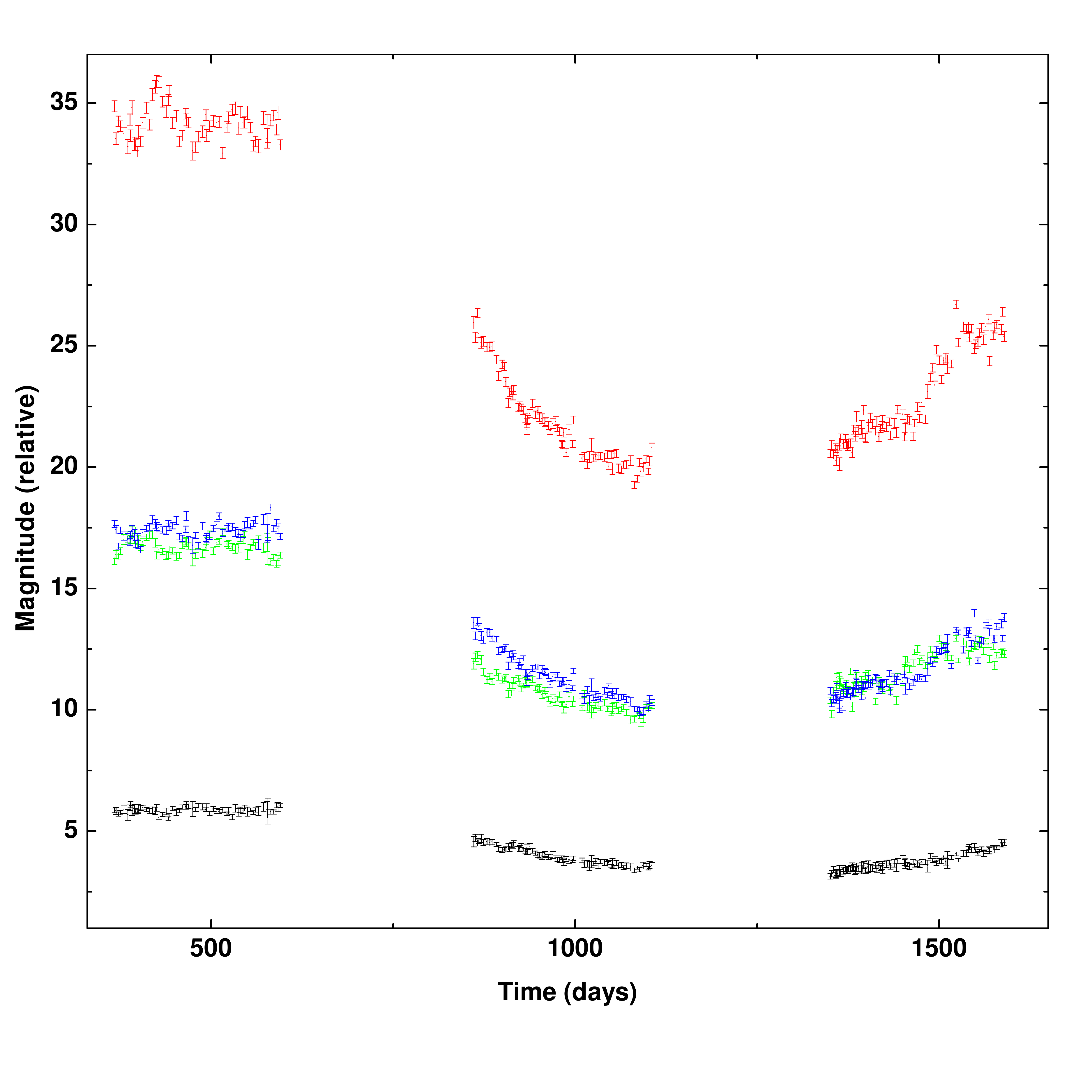}
\caption{Magnitudes (log flux) of four images of the quasar B1608+656
are plotted vs time, with an arbitrary overall zeropoint.  
}
\label{fig:1608}
\end{figure}

The quasar B1608+656 (lightcurves shown in Fig.~\ref{fig:1608}) is 
an example of a challenging system with large data gaps, relatively small 
intrinsic variability, and significant microlensing, all of which make it 
hard to estimate the time delays of its images. While we have successfully 
derived the time delays between all the images, including the cases not 
presented in the literature, the error bars are relatively large. This 
is in part due to the featureless light curves (especially the bottom 
curve, D in Table~\ref{tab:actual},  which is almost flat) and also due 
to the fact that our errors are marginalized over other parameters.  
For example, fitting the nugget term increases the errors by at least a 
factor of two while its presence is relatively unimportant for this system. 
We find that the probability distributions for some of the time delays in 
the DRW case have a smaller secondary peak, so comparison 
with the Matern results is useful to ensure robustness.

%%%%%%%%%%%%%%%%%%%%%%%%%%%

%%%%%%%%%%%%%%%%%%%%%%%%%%%%%%%%%%%%%%%%%%%%%%%%%%%%%%%%%%%%    
\section{Conclusions} \label{sec:concl}

Accurate estimation of strong lensing time delays is an essential 
element in the use of time delay distances as a novel cosmological probe.  
The complementarity, substantially geometric nature, and disjoint 
systematics of this technique make its use a goal worth striving for. 

We have explored Gaussian processes as a regression method that is 
effectively model independent and we demonstrated robust results for both 
blind mock data and actual literature data, in many cases reducing the 
uncertainties of the time delay estimations.  Noisy data, gaps in the 
observations, and extrinsic microlensing variations can all be handled 
by the method.  

Robustness arises not just from the technique itself, but the ability to 
use multiple lightcurves simultaneously, and test results against different 
combinations.  Several possibilities exist for further improvement.  For 
example one 
could weight the estimations derived from different combination of curves or 
one could remove unnecessary hyperparameters to reduce estimation uncertainty 
while checking that the best fit does not shift.  

Future data challenges will 
provide an opportunity to further develop the technique, providing 
important training and assessment of the reconstruction method.  
And of course one could 
obtain better real data.  Forthcoming surveys will find many more suitable 
lensing systems, allowing choice of the cleanest or best observed (with 
low photometric uncertainties, better cadence with fewer gaps, etc.).  

While time delay estimation is just one element in the development of 
strong lensing distances as a new cosmological probe, its improvement is 
key to this promising technique for mapping the Universe. 
Future work includes applying our GP reconstruction method to studies of 
lensed supernovae or other variable sources.

%%%%%%%%%%%%%%%%%%%%%%%%%%%%%%%%%%%%%%%%%%%%%%%%%%%%%%%%%%%%    

\acknowledgments 

We thank Chris Fassnacht for providing lightcurve data, Chris Kochanek 
for discussions on the DRW approach to intrinsic lightcurves, and Arman 
Shafieloo for discussions on GP code methodology.  AH acknowledges the 
Berkeley Center for Cosmological Physics for hospitality.  
This work has been supported by World Class University grant 
R32-2009-000-10130-0 through the National Research Foundation, Ministry 
of Education, Science and Technology of Korea and the Director, 
Office of Science, Office of High Energy Physics, of the U.S.\ Department 
of Energy under Contract No.\ DE-AC02-05CH11231.

%%%%%%%%%%%%%%%%%%%%%%%%%%%%%%%%%%%%%%%%%%%%%%%%%%%%%%%%%%%%


\begin{thebibliography}{}

\bibitem{oguri07} 
M. Oguri, ApJ 660, 1 (2007) [arXiv:astro-ph/0609694] 

\bibitem{suyu0804} 
S.H. Suyu, P.J. Marshall, R.D. Blandford, C.D. Fassnacht, L.V.E. Koopmans, 
J.P. McKean, T. Treu, ApJ 691, 277 (2009) [arXiv:0804.2827] 

\bibitem{cosmograil} 
F. Courbin et al, Astron. Astrophys. 536, A53 (2011) [arXiv:1009.1473] 

\bibitem{suyu0910} 
S.H. Suyu, P.J. Marshall, M.W. Auger, S. Hilbert, R.D. Blandford, 
L.V.E. Koopmans, C.D. Fassnacht, T. Treu, ApJ 711, 201 (2010) 
[arXiv:0910.2773] 

\bibitem{fadely} 
R. Fadely, C.R. Keeton, R. Nakajima, G.M. Bernstein, ApJ 711, 246 (2010) 
[arXiv:0909.1807] 

\bibitem{lin04} 
E.V. Linder, Phys. Rev. D 70, 043534 (2004) [arXiv:astro-ph/0401433] 

\bibitem{lin11} 
E.V. Linder, Phys. Rev. D 84, 123529 (2011) [arXiv:1109.2592] 

\bibitem{tewes} 
M. Tewes, F. Courbin, G. Meylan, arXiv:1208.5598

\bibitem{12086009} 
M. Tewes et al, arXiv:1208.6009 

\bibitem{12086010} 
S.H. Suyu et al, ApJ 766, 70 (2013) [arXiv:1208.6010] 

\bibitem{press} 
W.H. Press, G.B. Rybicki, J.N. Hewitt, ApJ 385, 404 (1992) 

\bibitem{des} 
Dark Energy Survey Collaboration, arXiv:astro-ph/0510346\\ 
\url{http://www.darkenergysurvey.org} 

\bibitem{kids} 
J.T.A. de Jong, G.A. Verdoes Kleijn, K.H. Kuijken, E.A. Valentin, 
Experimental Astron. 35, 25 (2013) [arXiv:1206.1254]\\ 
\url{http://kids.strw.leidenuniv.nl} 

\bibitem{panstarrs} 
N. Kaiser et al, Proc. SPIE 7733, 77330E (2010)\\ 
\url{http://pan-starrs.ifa.hawaii.edu} 

\bibitem{lsst} 
LSST Dark Energy Science Collaboration, arXiv:1211.0310\\ 
\url{http://www.lsst.org/lsst} 

\bibitem{koopmans} 
L.V.E. Koopmans, A.G. de Bruyn, E. Xanthopoulos, C.D. Fassnacht, Astr. 
Astroph. 356, 391 (2000) [arXiv:astro-ph/0001533] 

\bibitem{kimgp} 
A.G. Kim et al, ApJ 766, 84 (2013) [arXiv:1302.2925] 

\bibitem{grailweb} 
%\url{http://obswww.unige.ch/~tewes/cosmograil/public/lightcurves.php} 
\url{http://www.cosmograil.org} 

\bibitem{gpml}
C. E. Rasmussen \& C. K. I. Williams, Gaussian Processes for Machine 
Learning, MIT Press (2006) \\ 
\url{www.GaussianProcess.org/gpml}

\bibitem{drw1} 
B.C. Kelly, J. Bechtold, A. Siemiginowska, ApJ 698, 895 (2009) 
[erratum: ApJ 732, 128 (2011)] 
[arXiv:0903.5315] 

\bibitem{drw2} 
Y. Zu, C.S. Kochanek, B.M. Peterson, ApJ 735, 80 (2011) [arXiv:1008.0641] 

\bibitem{drw3} 
Y. Zu, C.S. Kochanek, S. Kozlowski, A. Udalski, ApJ 765, 106 (2013) 
[arXiv:1202.3783] 

\bibitem{drw4} 
R. Andrae, D-W. Kim, C.A.L. Bailer-Jones, arXiv:1304.2863 

\bibitem{minuit}
\url{http://seal.web.cern.ch/seal/work-packages/mathlibs/minuit/home.html} 

\bibitem{cosmomc} 
A. Lewis, S. Bridle, Phys. Rev. D 66, 103511 (2002) [arXiv:astro-ph/0205436]\\ 
\url{http://cosmologist.info/cosmomc}

%\cite{Vuissoz:2008qa}
\bibitem{Vuissoz:2008qa} 
%  C.~Vuissoz, F.~Courbin, D.~Sluse, G.~Meylan, V.~Chantry, E.~Eulaers, C.~Morgan and M.~E.~Eyler {\it et al.},
  %``COSMOGRAIL: the COSmological MOnitoring of GRAvItational Lenses VII. Time delays and the Hubble constant from WFI J2033-4723,''
C. Vuissoz et al, 
  Astron.\ Astrophys.\  { 488}, 481 (2008)
  [arXiv:0803.4015] 
  %%CITATION = ARXIV:0803.4015;%%

%\cite{Kochanek:2005ge}
\bibitem{Kochanek:2005ge} 
  C.~S.~Kochanek, N.~D.~Morgan, E.~E.~Falco, B.~A.~McLeod, J.~N.~Winn, J.~Dembicky, and B.~Ketzeback,
  %``The Time delays of gravitational lens HE0435-1223: An Early-type galaxy with a rising rotation curve,''
  Astrophys.\ J.\  { 640}, 47 (2006)
  [arXiv:astro-ph/0508070] 
  %%CITATION = ASTRO-PH/0508070;%%
  
  %\cite{Fassnacht:2002df}
\bibitem{Fassnacht:2002df} 
  C.~D.~Fassnacht, E.~Xanthopoulos, L.~V.~E.~Koopmans, and D.~Rusin,
  %``A Determination of H(O) with the class gravitational lens B1608+656. 3. A Significant improvement in the precision of the time delay measurements,''
  Astrophys.\ J.\  { 581}, 823 (2002)
  [arXiv:astro-ph/0208420] 
  %%CITATION = ASTRO-PH/0208420;%%


\end{thebibliography}
\end{document}